\documentclass[aps,prc,superscriptaddress]{revtex4}

\usepackage{graphicx}

\begin{document}

\title{ 
Effect of Shell Structure on Fission Isoscaling  
      }

\author{M. Veselsky}
\email{fyzimarv@savba.sk}
\affiliation{Institute of Physics of the Slovak Academy of 
Sciences, Dubravska 9, Bratislava, Slovakia}
\affiliation{Cyclotron Institute, Texas A\&M University, College Station, 
TX 77843}

\author{G. A. Souliotis}

\affiliation{Cyclotron Institute, Texas A\&M University, College Station, 
TX 77843}

\author{M. Jandel}
\affiliation{Institute of Physics of the Slovak Academy of 
Sciences, Dubravska 9, Bratislava, Slovakia}

\date{\today}

\nopagebreak

\begin{abstract}
Recently, the fragment yield ratios were investigated in 
the fission of $^{238,233}$U targets induced by 14 MeV neutrons \cite{FisIso}. 
The effect of the deformed shell structure around N=64 on the dynamics 
of scission was suggested as an explanation of the observed breakdown 
of the isoscaling behavior. As an alternative explanation, the effect 
of binding energy of final fragments, in particular the neutron shell 
closure of the complementary fragment around N=82, was suggested by 
Friedman \cite{Friedman}. When applying that concept consistently, the 
observed two breakdowns of the isoscaling behavior around N=62 and N=80 signal 
the effect of two shell closures ( N=82 and N=64 ) on the dynamics of scission. 
To separate the two effects, we examine the isoscaling plots obtained using 
fission fragment yield data from the spontaneous fission of the heavier 
nuclei $^{248,244}$Cm and find that the breakdown of isoscaling 
around N=64 persists even if it can not be related to shell closure 
of the complementary fragment and thus it manifests the effect 
of the deformed shell structure in the scission configuration.

\end{abstract}



\maketitle


Isoscaling phenomena were observed in the isotopic composition 
of light fragments originating from de-excitation of hot nuclei 
\cite{Tsang1} and also in heavy residue data in wide mass range 
\cite{GSHRIso}. Isoscaling is observed when a ratio 
of isotope yields from two reactions, differing in isospin 
asymmetry while other conditions are identical,  
exhibits an exponential dependence on N and  Z of the form 
$R_{21}(N,Z) \propto  \exp(\alpha N + \beta Z)$. 
In our recent study \cite{FisIso}, we presented  
the results of an isoscaling analysis 
of the fragment yield data from the fission induced by fast neutrons. 
The evaluated independent fragment yields from the fission 
of $^{233,238}$U targets induced by 14 MeV neutrons \cite{ENDF349} were used. 


    \begin{figure}[h]                                        

   \includegraphics[width=0.66\textwidth, height=0.5\textheight ]{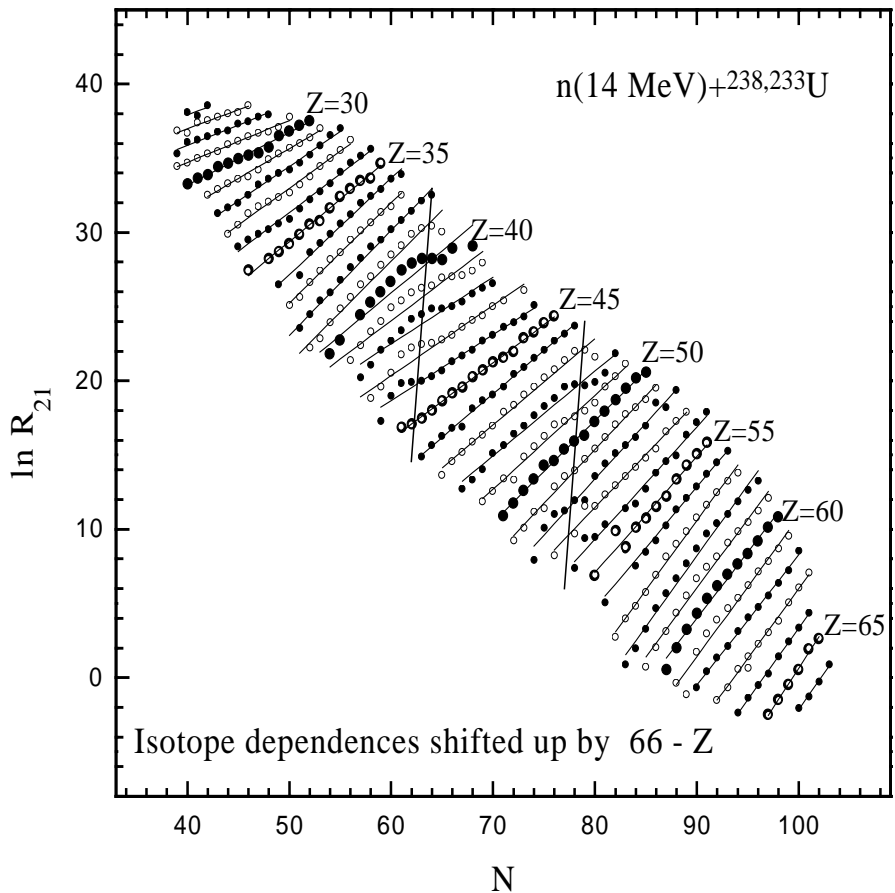}

    \caption{
Ratios of the fragment yields from the fission of $^{238,233}$U targets 
induced by 14 MeV neutrons \cite{ENDF349}. The data are shown as 
alternating solid and open circles. The labels apply to the larger 
symbols. The lines represent exponential fits. For clarity, the 
R$_{21}$ dependences are shifted from element to element by one unit. 
Nearly vertical lines mark major isoscaling breakdowns. 
           }
    \label{UFIscl}
    \end{figure}


The isoscaling behavior ( Fig. \ref{UFIscl} ) was typically observed for 
isotopic chains ranging from the most proton-rich to most neutron-rich ones. 
A breakdown of the isoscaling behavior was observed 
around N=64 which we suggested as a signal of 
the effect of the deformed shell structure on the dynamics of scission.  
This can be possibly related to the effect of the deformed neutron shell 
at N=64 with quadrupole deformation $\beta_{quad}$=0.6.  
According to the fission model of Wilkins et al. \cite{Wilkins}, 
this deformed shell closure plays a crucial role in the dynamics of 
scission. As an alternative explanation, a simple model based 
on the effect of the binding energy of final fragments was suggested 
by Friedman \cite{Friedman}. Assuming Boltzmann statistics with a single 
temperature for both fragments, the isoscaling yield ratio R$_{21}$ for 
a given fragment ($A_{1},Z_{1}$) is determined by the Boltzmann factor 
containing the difference of binding energies of the complementary fragments 
of the two fissioning systems.

\begin{equation}
R_{lh}(A_{1},Z_{1}) \propto  \exp(\frac{\Delta B_{ld}^{Z_{fiss}-Z_{1}}}{T}) 
                       \exp(\frac{\Delta B_{sh}^{Z_{fiss}-Z_{1}}}{T}) 
\label{eqn0}
\end{equation}

where $\Delta B_{ld}^{Z_{fiss}-Z_{1}}$ and $\Delta B_{sh}^{Z_{fiss}-Z_{1}}$ 
are differences 
of the liquid-drop and shell structure terms of the binding energies of the 
complementary fragments with charge $Z_{fiss}-Z_{1}$. 
The binding energies of the liquid drop model ( first factor ) lead 
to isoscaling behavior while the introduction of shell corrections 
around N=82 was shown to produce an effect around N=60-65, analogous 
to the effect observed in the data. The model uses shell corrections 
derived from the experimental ground state masses and thus corresponding 
to spherical nuclei. The deformations in the scission configuration 
are expected to be significant, with exception of mass splits with 
one fragment around Z=50 which is expected to be spherical \cite{Wilkins}. 
The isoscaling behavior obtained when assuming the liquid drop ground state 
binding energies will possibly not change qualitatively also when deformation 
will be introduced into the liquid drop model. The effect of deformation 
on the shell structure can again be expected to lead to isoscaling breakdowns 
corresponding to deformed shell closures. For the nuclei around N=82, 
the deformed shell closures are predicted 
for N=80 and N=88 \cite{Wilkins}, which may in principle lead to similar 
effect on isoscaling as the spherical neutron shell N=82. It is worthwile 
to note that, within the model of Friedman \cite{Friedman}, the neutron shell 
N=50 should result in an effect of similar strength in the region around N=94. 
However, such effect can not be identified unambigiously. On the other hand, 
the observed breakdown of isoscaling around N=78-80 can be, within the model 
of Friedman \cite{Friedman}, explained by the existence of shell closure 
around N=62-69, where no spherical shell is predicted, which again points 
to the influence of deformed shell closure around N=64. Thus, the observed 
behavior supports the presence of the deformed shell around N=64.  

\begin{figure}[h]                                        

   \includegraphics[width=0.66\textwidth, height=0.5\textheight ]{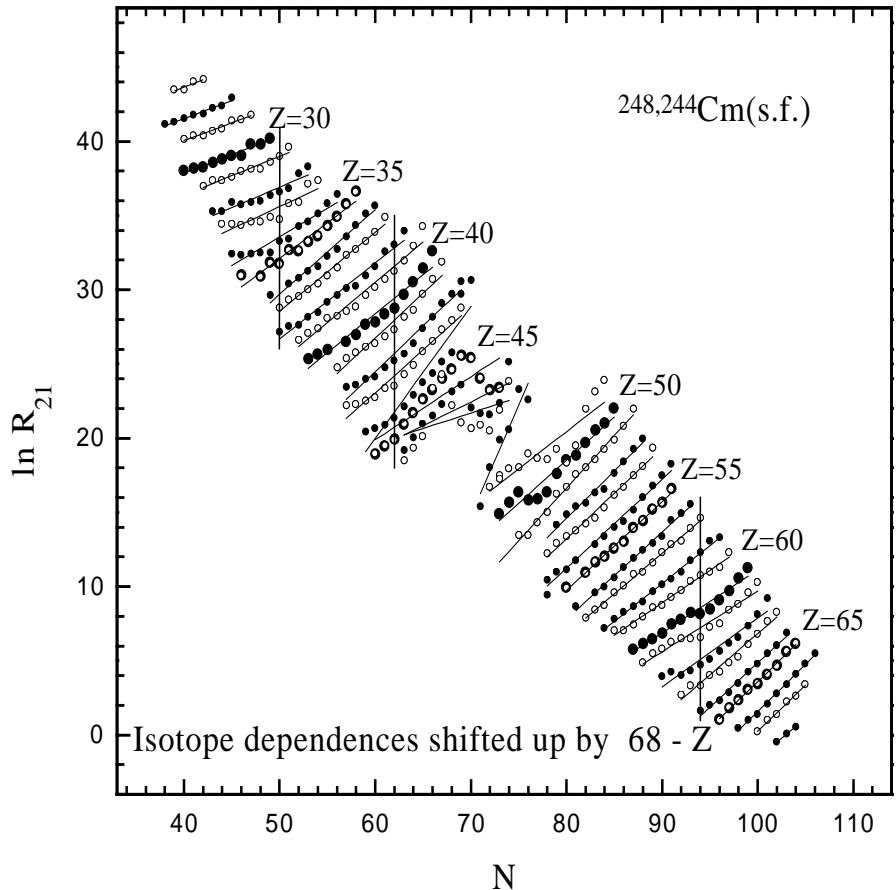}

    \caption{
Ratios of the fragment yields 
from the spontaneous fission of $^{248,244}$Cm \cite{ENDF349}. 
The data are shown as 
alternating solid and open circles. The labels apply to the larger 
symbols. The lines represent exponential fits. For clarity, the 
R$_{21}$ dependences are shifted from element to element by one unit. 
Vertical lines mark major isoscaling breakdowns. 
           }
    \label{CmFIscl}
    \end{figure}

As already mentioned, the model framework of Friedman \cite{Friedman} is based 
on the assumption that the probability of a given binary fission channel is 
determined by a Boltzmann factor depending on the sum of binding energies 
of two complementary fragments in the scission configuration. The observed data 
represent the inclusive yields corresponding to various multiplicities 
of neutrons emitted during fission. The neutron multiplicity distributions 
are rather wide, even for spontaneous fission \cite{CfNMult,PuNMult} and 
thus the concept of a unique complementary fragment is approximate. 
The next logical step in the analysis of influence of complementary fragments 
would be to examine the isoscaling behavior of fissioning systems with 
different mass. The systematic fission data for nuclei heavier than uranium 
are rather scarce, one nevertheless can find a good candidate in evaluated 
independent fragment yield data from spontaneous fission of $^{248,244}$Cm 
\cite{ENDF349}. The spontaneous fission is a much 
colder process where additional phenomena such as barrier penetration 
and potential energy surface can play a crucial role, 
it is nevertheless interesting to explore it in the context of isoscaling. 
The isoscaling plot for the spontaneous fission of $^{248,244}$Cm 
is shown in Fig. \ref{CmFIscl}.


In the spontaneous fission of $^{248,244}$Cm, the overall isoscaling behavior 
can be observed from the most populated parts of the mass distribution 
towards asymmetric mass splits. 
The breakdowns featuring rather the slope change than the structures 
observed in the previous case are observed around N=50 and N=62. 
A strong effect is observed around N=68-70 ( with the complementary fragment 
around N=82 ), which nevertheless can be a sign of transition 
into the symmetric region around N=74-76 where the behavior 
is rather irregular. Furthermore the isoscaling behavior is quite regular 
up to the heaviest fragments ( with exception of the region around N=94 where 
a hint of isoscaling breakdown can be observed ). The overall behavior 
can not be explained using the model of Friedman \cite{Friedman} which is
logical when assuming that the process is cold and the validity of 
statistical picture is questionable. It is however remarkable 
that isoscaling breakdown ( slope change ) is again observed around N=62 
( along with N=50 ). This again possibly points to the role of shell structure, 
which may influence the potential energy surface and thus the probability 
of a given fission channel. While around N=50 the shell structure can 
be identified with a spherical fragment, for the region around N=62 
the influence of the deformed shell is the most plausible. 
The observed isoscaling behavior in the spontaneous fission of $^{248,244}$Cm, 
despite a more complex interpretation, nevertheless again appears 
to signal the role of the deformed shell closures ( in particluar 
around N=64 ) in the fission of actinides. Since the isoscaling 
studies of fission enable to observe details of scission dynamics 
it is of interest to study the isoscaling behavior within the advanced 
models of fission, possibly in terms of transport theory, as indicated in 
\cite{FisIso}.

\bibliography{cmfisiso.bib}



\end{document}